\documentclass{ws-book975x65}
\usepackage{ws-book-har}           
\title{Advances in  Wave Turbulence}              
\makeindex
\begin{document}
\def\eg{{\it e.g.}}
\def\etal{{\it et al.}\ }
\def\ie{{\it i.e.}\ }
\def\vs{{\it vs}\ }
\def\rms{{\it r.m.s. }}
\def\uu{{\bf u}}
\def\xx{{\bf x}}
\def\kk{{\bf k}}
\def\pp{{\bf p}}
\def\qq{{\bf q}}
\def\AA{{\bf A}}
\def\aa{{\bf a}}
\def\HH{{\cal H}}
\def\pa{\parallel} 
\def\kpe{k_\perp}
\def\kpa{k_\parallel}
\def\ppe{p_\perp}
\def\ppa{p_\parallel}
\def\qpe{q_\perp}
\def\qpa{q_\parallel}
\def\bB{{\bf b}}
\def\bv{{\bf v}}
\newfam\msbmfam
\font\tenmsbm=msbm10 \textfont\msbmfam=\tenmsbm
\def\msbm{\fam\msbmfam\tenmsbm}
\def\RR{{\msbm R}}
\def\be{\begin{equation}}
\def\ee{\end{equation}}
\def\ba{\begin{eqnarray}}
\def\ea{\end{eqnarray}}

\tableofcontents
\setcounter{page}{1}

\bibliographystyle{ws-book-har}    
\bibliography{ws-book-sample}      


\chapter[Wave turbulence in magnetohydrodynamics]
{Wave turbulence in magnetohydrodynamics (MHD)}

This chapter reviews the recent progress made mainly during the last two decades on wave turbulence in 
magnetized plasmas (MHD, Hall MHD and electron MHD) in the incompressible and compressible cases. 
The emphasis is made on homogeneous and anisotropic turbulence which usually provides the best 
theoretical framework to investigate space and laboratory plasmas. The interplanetary medium and the 
solar atmosphere are presented as two examples of media where anisotropic wave turbulence is relevant. 
The most important results of wave turbulence are reported and discussed in the context of space and 
simulated magnetized plasmas. Important issues and possible spurious interpretations are eventually 
discussed. 

\section{Introduction}

Wave turbulence is the study of the long-time statistical behavior of a sea of weakly nonlinear 
dispersive waves \citep{ZLF,Naza11}. 
The energy transfer between waves occurs mostly among resonant sets 
of waves and the resulting energy distribution, far from a thermodynamic equilibrium, is 
characterized by a wide power law spectrum and a high Reynolds number. This range of 
wavenumbers -- the inertial range -- is generally localized between large scales at which energy 
is injected in the system (sources) and small scales at which waves break or dissipate (sinks). 
Pioneering works on wave turbulence date back to the sixties when it was established that the 
stochastic initial value problem for weakly coupled wave systems has a natural asymptotic 
closure induced by the dispersive nature of the waves and the large separation of linear and 
nonlinear time scales \citep{BS66,BN67,BN69}. In the meantime, \citeauthor{Zakh66} (1966) 
showed that the wave kinetic equations derived from the wave turbulence analysis (with a 
Gaussian Ansatz applied to the four-point correlations of the wave amplitude) have exact 
equilibrium solutions which are the thermodynamic zero flux solutions but also -- and more 
importantly -- finite flux solutions which describe the transfer of conserved quantities between 
sources and sinks. The solutions, first published for isotropic turbulence \citep{Zakh65,Zakh66} 
were then extended to anisotropic turbulence \citep{kuznetsov72}.  

Wave turbulence is a very common natural phenomenon with applications, for example, in capillary 
waves \citep{Kolmakov,Abdurakhimov}, gravity waves \citep{Falcon}, superfluid helium and processes 
of Bose-Einstein condensation \citep{Kolmakov95,lvov03}, nonlinear optics \citep{Dyachenko}, inertial 
waves \citep{Galtier2003,morize} or Alfv\'en waves \citep{Galtier2000,kuznetsov01,chandran}. 
The most important difference between plasmas and incompressible neutral fluids is the plethora 
of linear waves supported by the former. The direct consequence is that in weakly nonlinear plasmas 
the fundamental entities are waves rather than the eddies of strong turbulence \citep{K41,Krommes}. 
In the situation discussed here -- magnetized plasmas seen as fluids -- wave and strong turbulence 
may coexist and therefore both waves and eddies have in fact an impact on the nonlinear dynamics which 
is strongly anisotropic. Anisotropic turbulence is particularly well observed in space 
plasmas since a magnetic field is often present on the largest scale of the system, like in the inner 
interplanetary medium where the magnetic field lines form an Archimedean spiral near the equatorial 
plane \citep{Goldstein99}, at the solar surface where coronal loops and open magnetic flux tubes 
are found \citep{Cranmer07} or in planetary magnetospheres where shocks and discontinuities 
are measured \citep{Sahraoui06}.

In the present chapter, a review is made on wave turbulence in magnetized plasmas. A plasma is a 
gaseous state of matter in which the atoms or molecules are strongly ionized. Mutual electromagnetic 
forces between the ions and the free electrons are then playing dominant roles which adds to the 
complexity as compared to the situation in neutral atomic or molecular gases. In the situation discussed 
here the plasmas are described by the magnetohydrodynamics (MHD) approximation in the incompressible 
or compressible case. The role played by the Hall term is discussed through the Hall MHD description 
as well as its small-scale limit of electron MHD. 

The structure of the chapter is as follows. 
Physical motivations for developing wave turbulence theories are given in Section \ref{2} where we first 
describe multiscale solar wind turbulence, and then present the coronal heating problem. Section \ref{2bis} 
emphasizes the differences between strong and wave turbulence in MHD and the path followed historically 
by researches to finally obtain the MHD wave turbulence theories. In Section \ref{3}, the wave turbulence 
formalism is exposed with the basic ideas to derive the wave kinetic equations. Section \ref{4} deals with the 
results obtained under different approximations (MHD, Hall MHD and electron MHD) in the incompressible 
or compressible case. Finally we conclude with a general discussion in the last Section.

\section{Waves and turbulence in space plasmas}
\label{2}

Waves and turbulence are ubiquitous in astrophysical plasmas. Their signatures are found in 
the Earth's magnetosphere \citep{Sahraoui06}, the solar corona \citep{chae}, the solar wind 
\citep{bruno} or the interstellar medium \citep{elmegreen,scalo}. These regions are 
characterized by extremely large (magnetic) Reynolds numbers, up to $10^{13}$, with a range 
of available scales from $10^{18}$m to a few meters. 

\subsection{Interplanetary medium}

Extensive investigations are made in the interplanetary medium 
(and in the Earth's magnetosphere which is not the subject discussed here) where many 
{\it in situ} space-crafts measurements are available. The solar wind plasma is found to be in 
a highly turbulent state with magnetic and velocity fluctuations detected from $10^{-6}$Hz up 
to several hundred Hz \citep{Coleman,Roberts87,Leamon98,Smith06}. The turbulent state of 
the solar wind was first suggested in 1968 \citep{Coleman} when a power law behavior was 
reported for energy spectra with spectral indices lying between $-1$ and $-2$ (with the use 
of the Taylor ``frozen-in flow'' hypothesis). More precise measurements revealed that the 
spectral index at low frequency ($<1$Hz) is often about $-1.7$ which is closer to the 
Kolmogorov prediction \citep{K41} for neutral fluids ($-5/3$) rather than the Iroshnikov-Kraichnan 
prediction \citep{iro,Kraichnan65} for magnetized fluids ($-3/2$). 
Alfv\'en waves are also well observed since 1971 \citep{Belcher71} with a strong 
domination of antisunward propagative waves at short heliocentric distances (less than 1 AU). 
Since pure (plane) Alfv\'en waves are exact solutions of the ideal incompressible MHD equations 
\citep[see e.g.,][]{pouquet93}, nonlinear interactions should be suppressed if only one type of 
waves is present. Therefore sunward Alfv\'en waves, although subdominant, play an important 
role in the internal solar wind dynamics. 

The variance analysis of the magnetic field components and of its magnitude shows clearly that the 
magnetic field vector of the (polar) solar wind has a varying direction but with only a weak variation 
in magnitude \citep{forsyth96}. Typical values give a normalized variance of the field 
magnitude smaller than $10$\% whereas for the components it can be as large as $50$\%. In 
these respects, the inner interplanetary magnetic field may be seen as a vector lying approximately 
around an Archimedean spiral direction with only weak magnitude variations \citep{Barnes}. 
Solar wind anisotropy with more power perpendicular to the mean magnetic field than that parallel to it, 
is pointed out by data analysis \citep{Klein93} that provides a ratio of power up to $30$. From single-point 
spacecraft measurements it is not possible to specify the exact three-dimensional form of 
the spectral tensor of the magnetic or velocity fluctuations. However, it is possible to show that the 
spacecraft-frame spectrum may depend on the angle between the local magnetic field direction and the flow 
direction \citep{horbury}. In the absence of such data, a quasi 
two-dimensional model was proposed \citep{Bieb} in which wave vectors are nearly perpendicular 
to the large-scale magnetic field. It is found that about $85\%$ of solar wind turbulence possesses 
a dominant 2D component. Additionally, solar wind anisotropies is detected through radio wave 
scintillations which reveal that density spectra close to the Sun are highly anisotropic with irregularities 
stretched out mainly along the radial direction \citep{Armstrong90}. More recently 

For frequencies larger than $1$Hz, a steepening of the magnetic fluctuation power law spectra is 
observed over more than two decades \citep{coroniti,Denskat83,Leamon98,Bale,Smith06} with a 
spectral index close to $-2.5$. This new inertial range seems to be characterized by a 
bias of the polarization suggesting that these fluctuations are likely to be right-hand polarized, 
outward propagating waves \citep{Golstein94}. Various indirect lines of evidence indicate that 
these waves propagate at large angles to the background magnetic field and that the power in 
fluctuations parallel to the background magnetic field is still less than the perpendicular one 
\citep{coroniti,Leamon98}. For these reasons, it is thought \citep{Stawicki01} that Alfv\'en -- left 
circularly polarized -- fluctuations are suppressed by proton cyclotron damping and that the high 
frequency power law spectra are likely to consist of whistler waves. This scenario is supported 
by multi-dimensional direct numerical simulations of compressible Hall MHD turbulence in the 
presence of an ambient field 
\citep{Ghosh96} where a steepening of the spectra was found on a narrow range of wavenumbers, 
and associated with the appearance of right circularly polarized fluctuations. This result has been 
confirmed numerically with a turbulent cascade (shell) model based on 3D Hall MHD 
in which a well extended steeper power law spectrum was found at scale smaller than the ion skin depth 
\citep{Galtier07}. (Note that in this cascade model no mean magnetic field is assumed.) However, the 
exact origin of the change of statistical behavior is still under debate \citep{Marko08}: for example, an 
origin from compressible effects is possible in the context of Hall MHD \citep{Alexandrova}; a gyrokinetic 
description is also proposed in which kinetic Alfv\'en waves play a central role \citep{Howes08}. 

The solar wind plasma is currently the subject of a new extensive research around the origin of the 
spectral break observed in the magnetic fluctuations accompanied by an the absence of intermittency 
\citep{kiyani}. We will see that wave turbulence may 
have a central role in the sense that it is a useful point of departure for understanding the detailed 
physics of solar wind turbulence. In particular, it gives strong results in regards to the possible 
multiscale behavior of magnetized plasmas as well as the intensity of the anisotropic transfer 
between modes.

\subsection{Solar atmosphere}

\begin{figure}[t]
\centerline{\psfig{file=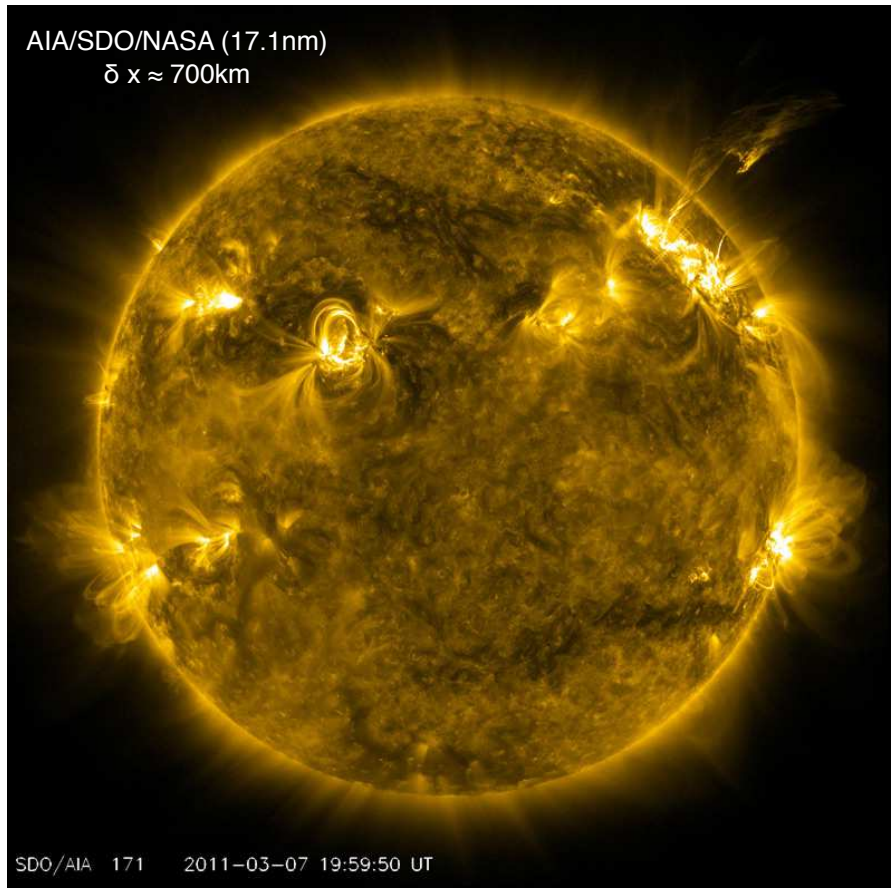,width=12cm}}
\caption{Solar corona seen at $17.1$nm by the AIA instrument onboard SDO (Solar Dynamics Observatory) 
-- NASA credits. The spatial resolution of the picture is about $700$km.}
\label{fig1}
\end{figure}

Although it is not easy to measure directly the coronal magnetic field, it is now commonly accepted 
that the structure of the low solar corona is mainly due to the magnetic field \citep{Aschwanden}. 
The high level of coronal activity appears through a perpetual impulsive reorganization of magnetic 
structures over a large range of scales, from about $10^5$ km until the limit of resolution, about one 
arcsec ($<726$~km). 
The origin of the coronal reorganization is currently widely studied in solar physics. Information about 
the solar corona comes from spacecraft missions like SoHO/ESA or TRACE/NASA launched in the 1990s, 
or from more recent spacecrafts STEREO/NASA, Hinode/JAXA or SDO/NASA (see Fig. \ref{fig1}). 
The most recent observations reveal that coronal loops 
are not yet resolved transversely and have to be seen as tubes made of a set of strands which radiate 
alternatively. In fact, it is very likely that structures at much smaller scales exist but have not yet been 
detected \citep[see e.g.,][]{warren}. 

Observations in UV and X-ray show a solar corona extremely hot with temperatures exceeding 
$10^6$K -- close to hundred times the solar surface temperature. These coronal temperatures are 
highly inhomogeneous: in the quiet corona much of the plasma lies near $1$--$2 \times 10^6$K and 
$1$--$8 \times 10^6$K in active regions. Then, one of the major questions in solar physics concerns 
the origin of such high values of coronal temperature. The energy available in the photosphere -- 
characterized by granules -- is clearly sufficient to supply the total coronal losses \citep{priest} 
which is about $10^4 {\rm J \, m}^{-2} {\rm s}^{-1}$ for active regions and about one or two 
orders of magnitude smaller for the quiet corona and coronal holes where open magnetic field lines 
emerge. The main issue is thus to understand how the available photospheric energy is transferred 
and accumulated in the solar corona, and by what processes it is dissipated. 

In active region loops, analyses made by spectrometers show that the plasma velocity can reach 
values up to $50$ km/s \citep{brekke}. The highly dynamical nature of some coronal loops is also 
pointed out by non-thermal velocities reaching sometimes $50$ km/s as it was revealed for example 
by SoHO/ESA \citep{chae}. These observations give also evidences that the line broadening is 
due to motions which are still not resolved neither in space, with scales smaller than the diameter of 
coronal loops, nor in time, with timescales shorter than the exposure time of the order of few seconds. 
These velocity measurements are very often interpreted as a signature of MHD turbulence where 
small scales are produced naturally {\it via} a nonlinear cascade of energy. In the light of the most 
recent observations, it seems fundamental to study, both theoretically and numerically, the impact of 
small-scale phenomena on the coronal heating. Note that recent Hinode/JAXA and SDO/NASA pictures 
seem to show a magnetic field controlled by plasma turbulence at all scales in which Alfv\'en waves are 
omnipresent \citep[see e.g.,][]{Doschek,Nishizuka,cargill}. Thus, the turbulent activity of the corona is one 
of the key issues to understand the heating processes. 

In the framework of turbulence, the energy supplied by the photospheric motions and transported by 
Alfv\'en waves through the corona is transferred towards smaller and smaller scales by nonlinear 
coupling between modes (the so-called energy cascade) until dissipative scales are reached from
which the energy is converted into heating. The main coronal structures considered in such a 
scenario are the magnetic loops which cover the solar surface. 
Each loop is basically an anisotropic bipolar structure anchored in the photosphere. It forms a tube 
of magnetic fields in which the dense and hot matter is confined. Because a strong guiding magnetic 
field ($\mathbf{B_0}$) is present, the nonlinear cascade that occurs is strongly anisotropic with small 
scales mainly developed in the $\mathbf{B_0}$ transverse planes. Most of the models published deals
with isotropic MHD turbulence \citep[see e.g.,][]{hendrix} and it is only very recently that anisotropy has 
been included in turbulent heating models \citep{buchlin}. 

The latest observations show that waves and turbulence are among the main ingredients of the solar 
coronal activity. Weak MHD turbulence is now invoked has a possible regime for some coronal loops 
since a very small ratio is found between the fluctuating magnetic field and the axial component
\citep{Rappazzo, Rappazzo2}. Inspired by the observations and by recent direct numerical simulations 
of 3D MHD turbulence \citep{Bigot08b}, an analytical model of coronal structures has been proposed 
\citep{Bigot08c} where the heating is seen as the end product of a wave turbulent cascade. Surprisingly, 
the heating rate found is non negligible and may explain the observational predictions. 

The coronal heating problem also concerns the regions where the fast solar wind is produced, 
\ie the coronal holes \citep{hollweg,Cranmer07}.
Observations seem to show that the heating affects preferentially the ions in the direction perpendicular 
to the mean magnetic field. The electrons are much cooler than the ions, with temperatures generally 
less than or close to $10^6$K \citep[see \eg,][]{david}. Additionally, the heavy ions become hotter than 
the protons within a solar radius of the coronal base. Ion cyclotron waves could be the agent which heats 
the coronal ions and accelerates the fast wind. Naturally the question of the origin of these high frequency 
waves arises. Among different scenarios, turbulence appears to be a natural and efficient mechanism to 
produce ion cyclotron waves. In this case, the Alfv\'en waves launched at low altitude with frequencies in 
the MHD range, would develop a turbulent cascade to finally degenerate and produce ion cyclotron 
waves at much higher frequencies. In that context, the wave turbulence regime was considered in the 
weakly compressible MHD case at low-$\beta$ plasmas (where $\beta$ is the ratio between the 
thermal and magnetic pressure) in order to analyze the nonlinear three-wave interaction transfer  
to high frequency waves \citep{chandran}. The wave turbulence calculation shows -- in absence of slow 
magnetosonic waves -- that MHD turbulence provides a convincing explanation for the anisotropic ion heating.

\section{Turbulence and anisotropy}
\label{2bis}

This Section is first devoted to the comparison between wave and strong turbulence. In particular, we will 
see how the theoretical questions addressed at the end of the 20th century have led to the emergence of 
a large number of papers on wave turbulence in magnetized plasmas and to many efforts to characterize 
the fundamental role of anisotropy.

\subsection{Navier-Stokes turbulence}

Navier-Stokes turbulence is basically a strong turbulence problem in which it is impossible to perform a 
(non trivial and) consistent linearization of the equations against a stationary homogeneous background. 
We remind that wave turbulence demands the existence of linear (dispersive) propagative waves
as well as a large separation of linear and nonlinear (eddy-turnover) time scales \citep[see \eg,][]{BN69}.
In his third 1941 turbulence paper, \citet{K41} found that an exact and nontrivial relation may be derived 
from Navier-Stokes equations for the third-order longitudinal structure function \citep{K41}: it is the so-called 
$4/5$s law. Because of the rarity of such results, the Kolmogorov's $4/5$s law is considered as one of the 
most important results in turbulence \citep{frisch}. Basically, this law makes the following link in the 3D physical 
space between a two-point measurement, separated by a distance ${\bf r}$, and the distance 
itself\footnote{The Kolmogorov $4/5$s law can be written in an equivalent way as a $4/3$s law \citep{antonia,ras},
namely: $- (4 /3) \varepsilon^v r = \langle ({v_L^{\prime}} - v_L) \sum_i ({v_i^{\prime}} - v_i)^2 \rangle$.}
\be
- {4 \over 5} \varepsilon^v r = \langle (v_L^{\prime} -v_L)^3 \rangle \, ,
\ee
where $\langle \rangle$ denotes an ensemble average, the longitudinal direction $L$ is the one along 
the vector separation ${\bf r}$, $v$ is the velocity and $\varepsilon^v$ is the mean (kinetic) energy 
injection or dissipation rate per unit mass. To obtain this exact relation, the assumptions of 
homogeneity and isotropy are made \citep{batch}. The former assumption is satisfied as long as we are 
at the heart of the fluid (far from the boundaries) and the latter is also satisfied if no external agent (like, 
for example, rotation or stratification) is present. 
Additionally, the long time limit is considered for which a stationary state is reached with a finite 
$\varepsilon^v$ and the infinite Reynolds number limit (\ie the viscosity $\nu \to 0$) is taken for which 
the mean energy dissipation rate per unit mass tends to a finite positive limit. Therefore, the exact prediction 
is valid in the asymptotic limit of a large inertial range. The Kolmogorov law is well supported by the 
experimental data \citep[see \eg,][]{frisch}. 

The $4/5$s law is a fundamental result used to develop heuristic spectral scaling laws like the 
famous -- but not exact -- 5/3-Kolmogorov energy spectrum. This point makes a fundamental difference 
with wave turbulence where the power law spectra found are exact solutions of the asymptotically 
exact wave turbulence equations. Nevertheless, the term "Kolmogorov theory" is often associated to 
the $-5/3$ spectrum since there exists a theory behind in the physical space.

\subsection{Incompressible MHD Turbulence}
\subsubsection{Strong turbulence}

The wave turbulence regime exists in incompressible MHD. The main reason is that Alfv\'en waves are 
linear solutions when a stationary homogeneous background magnetic field ${\bf B_0}$ is applied. This 
statement seems to be obvious but we will see that the problem is subtle and the existence of an Alfv\'en 
wave turbulence theory was the subject of many discussions basically because those waves are only 
pseudo-dispersive (\ie, the frequency $\omega$ is proportional to a wavenumber). 

The question of the existence of an exact relation for third order structure functions is naturally addressed 
for strong (without ${\bf B_0}$) MHD turbulence. The answer was given by Politano \& Pouquet only in 1998 
\citep[see also, ][]{chandra51} for incompressible MHD turbulence. The presence of the magnetic field 
and its coupling with the velocity field renders the problem more difficult and, in practice, we are dealing 
with a couple of equations. In this case, the possible formulation in 3D is the $4/3$'s law
\be
- {4 \over 3} \varepsilon^{\pm} r =
\langle ({z_L^{\prime}}^{\mp} - z_L^{\mp}) \sum_i ({z_i^{\prime}}^{\pm} - z_i^{\pm})^2 \rangle \, ,
\label{PP98}
\ee
where the direction $L$ is still the one along the vector separation ${\bf r}$, 
${\bf z^\pm} = {\bf v} \pm {\bf b}$ is the Els\"asser fields (with ${\bf b}$ normalized to a velocity field)
and $\varepsilon^\pm$ is the mean energy dissipation rate per unit mass associated to the Els\"asser 
energies. To obtain these exact relations, the assumptions of homogeneity and isotropy are still made, and 
the long time limit for which a stationary state is reached with a finite $\varepsilon^\pm$ is also considered. 
The infinite kinetic and magnetic Reynolds number limit ($\nu \to 0$ and the magnetic diffusivity $\eta \to 0$) 
for which the mean energy dissipation rates per unit mass have a finite positive limit is eventually taken. 
Therefore, the exact prediction is again valid, at first order, in a wide inertial range. This prediction is 
currently often used in the literature to analyze space plasma data and determine the local heating rate 
$\varepsilon^\pm$ in the solar wind \citep[see \eg,][]{sorriso,macbride}.

\subsubsection{Iroshnikov--Kraichnan spectrum}

\begin{figure}[b]
\centerline{\psfig{file=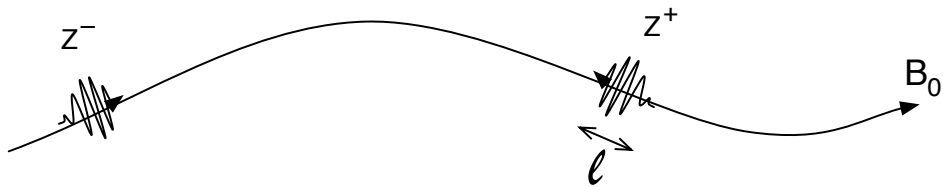,width=13cm}}
\caption{Alfv\'en wave packets propagating along a large scale magnetic field ${\bf B_0}$.}
\label{fig2}
\end{figure}

The isotropy assumption used to derive the 4/3's law is stronger for magnetized than neutral fluids since in 
most of the situations encountered in astrophysics a large scale magnetic field ${\bf B_0}$ is present which 
leads to anisotropy (see Section \ref{2}). Although this law is a fundamental result that may be used to develop 
a heuristic spectral scaling law, the role of $B_0$ has to be clarified. Indeed, we have now two time-scales: 
the eddy-turnover time and the Alfv\'en time. The former is similar to the eddy-turnover time in Navier-Stokes 
turbulence and may be associated to the distortion of wave packets (the basic entity in MHD), whereas the latter 
may be seen as the duration of interaction between two counter-propagating Alfv\'en wave packets (see 
Fig.\ref{fig2}).
During a collision, there is a deformation of the wave packets in such a way that energy is transferred 
mainly at smaller scales. The multiplicity of collisions leads to the formation of a well extended power 
law energy spectrum whose index lies between $-5/3$ (Kolmogorov prediction) and $-3/2$ 
(Iroshnikov-Kraichnan prediction) according to the phenomenology used, \ie with or without the 
Alfv\'en wave effect. Note the slightly different approaches followed by Iroshnikov and Kraichnan to 
derive the $-3/2$ spectrum. In the former case the presence of a strong magnetic field is explicitly assumed 
whereas it is not in the latter case where it is claimed that the small-scale fluctuations see the large-scales 
-- in the sub-inertial range -- as a spatially uniform magnetic field. However, they both assumed isotropy to 
finally proposed the so-called Iroshnikov-Kraichnan spectrum for MHD turbulence. 
It is important to note that the exact isotropic relation (\ref{PP98}) in physical space corresponds 
dimensionally to a $-5/3$ energy spectrum since it is cubic in $z$ (or in $v$ and $b$) and linear in $r$. 
It is therefore less justified to use the term "theory" for the Iroshnikov-Kraichnan spectrum than for the 
Kolmogorov one for neutral fluids.

\subsubsection{Breakdown of isotropy}
\label{resona}

The weakness of the Iroshnikov-Kraichnan phenomenology is the apparent contradiction between
the (in)direct presence of a strong uniform magnetic field and the assumption of isotropy. An important 
difference between neutral and magnetized fluids is the impossibility in the latter case to remove a 
large-scale (magnetic) field by a Galilean transform. The role of a uniform magnetic field has been widely 
discussed in the literature and, in particular, during the last three decades 
\citep[see \eg,][]{montgo1,sheba,Bill96,ngb1,verma}. At strong ${\bf B_0}$ intensity, one of the most 
clearly established results is the bi-dimensionalization of MHD turbulent flows with a strong reduction 
of nonlinear transfers along $\bf {B_0}$. The consequence is an energy concentration near the plane 
${\bf k} \cdot {\bf B_0} =0$, a result illustrated later on by direct numerical simulations in two and three 
space dimensions \citep{sheba}. 

The effects of a strong uniform magnetic field may be handled through an analysis of resonant triadic 
interactions \citep{sheba} between the wavevectors ($\kk,\pp,\qq$) which satisfy the vectorial relation 
\be
\kk = \pp + \qq \, ,
\ee 
whereas the associated wave frequencies satisfy the scalar relation
\be
\omega(\kk) = \omega(\pp) + \omega(\qq) \, .
\ee 
For incompressible MHD, the Alfv\'en frequency is 
\be
\omega(\kk)= \pm \kk \cdot {\bf B_0}= \pm \kpa B_0 \, ,
\ee
where $\pa$ defines the direction along ${\bf B_0}$ ($\perp$ will be the perpendicular direction to 
${\bf B_0}$ which is written in velocity unit). The solution of the three-wave resonance condition gives for 
example, $\qpa=0$, which implies a spectral transfer only in the perpendicular direction. 
For a strength of $B_0$ well above the \rms level of the kinetic and magnetic fluctuations, the nonlinear 
interactions of Alfv\'en wave packets may dominate the dynamics of the MHD flow leading to the regime of 
wave turbulence (see Section \ref{3}) where the energy transfer, stemming from three-wave resonant 
interactions, can only increase the perpendicular component of the wavevectors, while the nonlinear 
transfers are completely inhibited along ${\bf B_0}$. The end result is a strongly anisotropic flow.

\subsubsection{Emergence of anisotropic laws}

An important issue discussed in the literature is the relationship between perpendicular and parallel 
scales in anisotropic MHD turbulence \citep[see \eg,][]{higdon,gosri,boldyrev,sridhar10}. In order to take into account 
anisotropy, \citet{gosri} proposed a heuristic model based on the conjecture of a critical balance between 
the Alfv\'en and the eddy-turnover times, which are respectively 
\be
\tau_A \sim \ell_\parallel/B_0
\ee
and 
\be
\tau_{eddy}\sim \ell_\perp/u_{\ell} \, , 
\ee 
where $\ell_\parallel$ and $\ell_\perp$ are typical length scales parallel and perpendicular to ${\bf B_0}$.
The conjecture says that in the inertial range we have, $\tau_A = \tau_{eddy}$. The latter relation leads trivially to, 
$u_{\ell} \sim B_0 \ell_\perp / \ell_\parallel$. Following the Kolmogorov arguments, one ends up trivially with the 
Kolmogorov energy spectrum
\be
E(\kpe,\kpa) \sim \kpe^{-5/3} \kpa^{-1} \, , 
\label{5s3}
\ee
(where ${\bf k} \equiv$ (${\bf k}_\perp$, $\kpa$), $\kpe \equiv |{\bf k}_\perp|$,  
$\int \int E(\kpe,\kpa) d\kpe d\kpa = \int E(\kpe) d\kpe$), and with the non trivial anisotropic scaling law 
(with $u_{\ell}^2 / \tau_{eddy}=$ constant)
\be
\kpa \sim \kpe^{2/3} \, .
\label{aniso1}
\ee
This heuristic prediction means that anisotropy is stronger at smaller scales. 

A generalization of this result was proposed by \citet{Galtier2005b} in order to model MHD flows both in 
the wave and strong turbulence regimes, as well as for the transition between them. In this heuristic model, 
the constant time-scale ratio $\chi=\tau_A/\tau_{eddy}$ is not assumed to be necessarily equal to unity. 
The relaxation of the constraint ($\chi=1$) allows us to recover the anisotropic scaling law (\ref{aniso1}) which 
now includes $B_0$, 
\be
\kpa \sim \kpe^{2/3}/ B_0 \, ,
\label{aniso2}
\ee
and to find a universal prediction for the total energy spectrum 
\be
E(\kpe,\kpa) \sim \kpe^{-\alpha} \kpa^{-\beta} \, , 
\label{ab}
\ee
with
\be
3\alpha + 2 \beta =7 \, . 
\label{relab}
\ee
Note that a classical calculation with a transfer time $\tau = \tau_{eddy}^2 / \tau_A =  \tau_{eddy} / \chi$ 
leads trivially to the spectrum (\ref{5s3}): the choice of $\tau$ fixes irreversibly the spectrum. It is only when the 
ansatz (\ref{ab}) is introduced that the non trivial relation (\ref{relab}) emerges; this ansatz is believed to be weak 
since power laws are the natural solutions of the problem.  
According to direct numerical simulations \citep[see \eg,][]{cho2000,maron,ngb2,shaikh,Bigot08b}, the 
anisotropic scaling law between parallel and perpendicular scales (\ref{aniso1}) seems to be a robust 
result and an approximately constant ratio $\chi$, generally smaller than one, is found between the 
Alfv\'en and the eddy-turnover times. This sub-critical value of $\chi$ implies therefore a dynamics mainly 
driven by Alfv\'en waves interactions. Note that the presence of $B_0$ in relation (\ref{aniso2}) shows 
the convergence towards wave turbulence ($B_0 \to + \infty$, with respect to the fluctuations) 
for which the parallel transfer is totally frozen. 

The question of the spectral indices is still a challenging problem in anisotropic turbulence \citep{sagaut}. 
The main conclusion of \citet{Bigot08b} is that the difficulty to make the measurements is generally 
underestimated in a sense that the scaling prediction in $\kpe$ may change significantly 
when $E(\kpe,\kpa)$ is plotted at a given $\kpa$ instead of $E(\kpe)$: indeed, the slow mode  
$E(\kpe,\kpa=0)$ may play a singular role in the dynamics with a scaling in $\kpe$ different from 
the one given by the 3D modes $E(\kpe,\kpa>0)$.
This comment holds primarily for direct numerical simulations where technically it is currently possible to make 
this distinction, observations being still far from this possibility. This point will be further discussed in the last 
Section. Note finally that all these spectral predictions suffer from rigorous justifications and the word "theory" 
that we find very often in the literature is not justified at all. A breakthrough could be achieved if one could 
develop the equivalent of an exact $4/5$'s law for anisotropic MHD turbulence. Then, a dimensional 
derivation from this law could lead to an anisotropic spectral prediction and in some sense, for the first time, 
the possibility of having a theoretical link between strong and wave turbulence. Recently, an attempt has been 
made in that direction by using the idea of critical balance for third-order moments \citep{gal12}. Moreover, it 
was shown that the introduction of the dynamic alignment conjecture into the exact relation for third-order 
moments \citep{PP98} may give a $\kpe^{-3/2}$ spectrum \citep{boldyrev09}.

\subsection{Towards an Alfv\'en wave turbulence theory}

In view of the importance of anisotropy in natural magnetized plasma (see Section \ref{2}), 
\citet{sridhar} suggested that a plasma evolving in a medium permeated by a strong uniform 
magnetic field and in the regime of wave turbulence is characterized by four-wave nonlinear 
interactions. The essence of wave turbulence is the statistical study of large ensembles of weakly 
interacting waves via a systematic asymptotic expansion in powers of small nonlinearity. This 
technique leads finally to the exact derivation of wave kinetic equations for the invariants of the 
system like the energy spectrum (see Section \ref{3}). In MHD, it is the presence of a strong uniform 
magnetic field ${\bf B_0}$ that allows to introduce a small parameter in the system, namely the 
ratio between the fluctuating fields and $B_0$. 
The result found by Sridhar and Goldreich in 1994 implies that the asymptotic development has no 
solution at the main order and that we need to go to the next order to describe Alfv\'en wave turbulence. 
Several articles, using a phenomenology \citep[see \eg,][]{Mont95,verma} or a rigorous treatment \citep{ngb1}, 
were published to contest this conclusion and sustain the non trivial character of the three-wave interactions. 
In response, a detailed theory was finally given in 2000 \citep{Galtier2000,naza,Galtier2002} whose main 
prediction may be derived heuristically in few lines as follows. According to Fig. \ref{fig2} the main process 
which happens in Alfv\'en wave turbulence is the stochastic collisions of wavepackets. To find the transfer time 
and then the energy spectrum, first we shall evaluate the modification of a wavepacket produced by one collision. 
We have (for simplicity we only consider the balanced case for which $z^\pm_\ell \sim z_\ell \sim u_\ell \sim b_\ell$) 
\be
z_\ell(t+\tau_A) \sim z_\ell(t) + \tau_A {\partial z_\ell \over \partial t} \sim z_\ell(t) + \tau_A {z^2_\ell \over \ell_\perp} \, ,
\ee
where $\tau_A$ is the duration of one collision; in other words, after one collision the distortion of a wavepacket is 
$\Delta_1 z_\ell \sim \tau_A z^2_\ell / \ell_\perp$. This distortion is going to increase with time in such a way that after 
$N$ stochastic collisions the cumulative effect may be evaluated like a random walk
\be
\sum_{i=1}^N \Delta_i z_\ell \sim \tau_A {z^2_\ell \over \ell_\perp} \sqrt{t \over \tau_A} \, .
\ee
The transfer time $\tau_{tr}$ that we are looking for is the one for which the cumulative distortion is of the order 
of one, \ie of the order of the wavepacket itself
\be
z_\ell \sim \tau_A {z^2_\ell \over \ell_\perp} \sqrt{\tau_{tr} \over \tau_A} \, ,
\ee
then we obtain
\be
\tau_{tr} \sim {1 \over \tau_A} {\ell_\perp^2 \over z^2_\ell} \sim {\tau^2_{eddy} \over \tau_A} \, . 
\ee
A classical calculation, with $\varepsilon \sim z^2_\ell / \tau_{tr}$, leads finally to the energy spectrum 
\be
E(\kpe,\kpa) \sim \sqrt{\varepsilon B_0} \, \kpe^{-2} \kpa^{-1/2} \, ,
\ee
where $\kpa$ has to be seen as a parameter since no transfer along the parallel direction is expected
(see Section \ref{resona}). Note that this demonstration is the one traditionally used for deriving the 
Iroshnikov-Kraichnan spectrum, but since in this case isotropy is assumed a 
$E(k) \sim \sqrt{\varepsilon B_0} \, k^{-3/2}$ is predicted. 

First signatures that may be attributed to wave turbulence were found by \citet{Perez08} in 
numerical simulations of a reduced form of the MHD equations \citep{Galtier06b}. 
The detection of the wave turbulence regime from direct numerical simulations of MHD equations is still 
a difficult task but recent results have been obtained in which temporal, structural 
and spectral signatures are reported \citep{Bigot08a,Bigot08b,Bigot11}. Current efforts are also made to 
analyze the effects of other inviscid invariants, like the cross-helicity, on the scaling laws of wave 
turbulence \citep{LG03,chandran08}. It is worth noting that these works on imbalanced wave turbulence 
have been followed by (sometimes controversial) research investigations in the strong turbulence 
regime which is of great relevance for the solar wind \citep[see \eg][]{chandran09,beresnyak10,Pod10}.

\subsection{Wave turbulence in compressible MHD}

Most of the investigations devoted to wave turbulence refers to isotropic media where the well known 
conformal transform proposed by \citet{Zakh66} may be applied to find the so-called Kolmogorov-Zakharov 
spectra (see Section \ref{K41solution}). (Surprisingly a similar transform was used in the meantime by 
\citet{Kraichnan67} to investigate the problem of 2D turbulence.) The introduction of anisotropy in plasmas 
was studied to a smaller extent: the first example is given by magnetized ion-sound waves 
\citep{kuznetsov72}. The compressible  MHD case was analyzed later by \citet{kuznetsov01} for a situation 
where the plasma (thermal) pressure is small compared with the magnetic pressure (small $\beta$ limit). 
In this case, the main nonlinear interaction involving MHD waves is the scattering of a fast magneto-acoustic 
and Alfv\'en waves on slow magneto-acoustic waves. In this process, the fast and Alfv\'en waves act as 
high-frequency waves with respect to the slow waves. (To simplify the analysis other three-wave interaction 
processes that do not involve slow waves are neglected.) 

A variant of the wave turbulence analysis in compressible MHD was proposed by \citet{chandran} 
in the limit of a small $\beta$ in which the slow waves are neglected and a constant density is imposed. 
The other (mathematical) difference is that the Hamiltonian formalism was used in the former analysis
whereas an Eulerian description was employed in the latter case. Because the compressible regime is 
much more difficult to analyze than the incompressible one, simplifications have been made and to date, 
no general theory has been proposed for compressible MHD wave turbulence.

\subsection{Wave turbulence in Hall and electron MHD}

Modeling the physics of a plasma beyond the MHD approximation, namely on spatial scales shorter 
than the ion inertial length $d_i$ (but larger than the electron inertial length $d_e$) and time scales 
shorter than the ion cyclotron period $1/\omega_{ci}$, is a highly challenging problem even in the fluid 
approximation. (For kinetic models we can mention the gyrokinetic approximation useful for weakly 
collisional plasmas  in which time scales are supposed to be much larger than $1/\omega_{ci}$ 
\citep[see \eg,][]{scheko}.) In that context the electron MHD approximation \citep{Kingsep} is often 
used: in such a limit one assumes that ions do not have time to follow electrons and provide a static 
homogeneous background on which electrons move. The electron MHD approximation is particularly 
relevant in the context of collisionless magnetic reconnection where a diffusion region is developed 
with multiscale structures corresponding to ion and electron characteristic lengths 
\citep{Biskamp97,amitava,yamada}. 

An important issue in electron MHD turbulence is about the impact of whistler waves on the dynamics. 
\citet{Biskamp99} argued that although whistler wave propagation effects are non negligible in 
electron MHD turbulence, the local spectral energy transfer process is independent of the linear 
wave dispersion relation and the energy spectrum may be predicted by a Kolmogorov type argument. 
Direct numerical simulations were used to illustrate the theory but no mean magnetic field was introduced 
\citep{Biskamp96,ngb2}. \citet{dastgeer} investigated the turbulence regime in the presence of a moderate 
background magnetic field $B_0$ and provided convincing numerical evidence that turbulence is anisotropic. 
It was argued that although whistler waves may appear to play a negligible role in determining the spectral 
index, they are important in setting up an anisotropic turbulent cascade. The whistler wave turbulence 
regime was then investigated theoretically by \citet{Galtier03,Galtier05a} in the limit $B_0 \to + \infty$ (with 
respect to the fluctuations). It was shown that similarly to the MHD case, anisotropy is a central feature 
of such a turbulence. Attempts to find an anisotropic law for electron MHD was made by \citet{cho2004}
\citep[see also,][]{Galtier2005b} and a scaling relation in $\kpa \sim \kpe^{1/3}$ was found both heuristically 
and numerically. 

Hall MHD is an extension of the standard MHD where the ion inertia is retained in Ohm's law. It provides a 
multiscale description of magnetized plasmas from which both standard and electron MHD approximations 
may be recovered. Hall MHD is often used to understand, for example, the magnetic field evolution in neutron 
star crusts \citep{Goldreich92}, the turbulent dynamo \citep{Mininni03}, the formation of filaments \citep{passot}, 
the multiscale solar wind turbulence \citep{Ghosh96,krishan,Galtier06a,Galtier06aa}, or the dynamics of the 
magnetosheath \citep{Belmont01}.
Anisotropy in Hall MHD is clearly less understood than in MHD or electron MHD mainly because the 
numerical treatment is more limited since a wide range of scales is necessary to detect any multiscale 
effects. From a theoretical point of view, it is only recently that a wave turbulence theory has been 
achieved for the incompressible case \citep{Galtier06aa,Sahraoui07}. 
For such a turbulence, a global tendency towards anisotropy was found (with, however, a weaker 
effect at intermediate scales) with nonlinear transfers preferentially in the direction perpendicular 
to the external magnetic field ${\bf B_0}$. The energy spectrum is characterized by two inertial ranges, 
separated by a knee, which are exact solutions of the wave kinetic equations. The position of the knee 
corresponds to the scale where the Hall term becomes sub/dominant. To date, compressible Hall MHD is 
still an open problem in the regime of wave turbulence. (A first step was made by \citet{Sahraoui03} who 
found the Hamiltonian system.)

\section{Wave turbulence formalism}
\label{3}

\subsection{Wave amplitude equation}

Wave turbulence is the study of the long time statistical behavior of a sea of weakly nonlinear dispersive 
waves. It is described by wave kinetic equations. In this section we present the wave turbulence formalism 
which leads to these nonlinear equations. We shall use the inviscid model equation
\be
{\partial \uu(\xx,t) \over \partial t} = {\cal L}(\uu) + \varepsilon \, {\cal N}(\uu,\uu) \, ,
\label{eq1}
\ee
where $\uu$ is a stationary random vector, ${\cal L}$ is a linear operator which insures that waves 
are solutions of the linear problem, and ${\cal N}$ is a quadratic nonlinear operator (like for MHD-type 
fluids). The factor $\varepsilon$ is a small parameter ($0 < \varepsilon \ll 1$) which will be used for the 
weakly nonlinear expansion.  For all the applications considered here, the smallness of the nonlinearities 
is the result of the presence of a strong uniform magnetic field ${\bf B_0}$; the operator ${\cal L}$ is 
thus proportional to $B_0$. 

We introduce the 3D direct and inverse Fourier transforms 
\be
\uu(\xx,t) = \int_{\RR^3} \AA(\kk,t) \exp(i \kk \cdot \xx) d\kk \, , 
\ee
\be
\AA(\kk,t) = {1 \over (2\pi)^3} \int_{\RR^3} \uu(\xx,t) \exp(- i \kk \cdot \xx) d\xx \, .
\ee
Therefore, a Fourier transform of equation (\ref{eq1}) gives for the j-component 
\be
\left( {\partial \over \partial t} + i \omega(\kk) \right) A_j (\kk,t) =
\ee
$$ 
\varepsilon \int_{\RR^6} \HH_{jmn}^{\kk \pp \qq} A_m(\pp,t) A_n(\qq,t) \delta(\kk-\pp-\qq) d\pp d\qq \, ,
$$
where $\omega(\kk)=\omega_k$ is given by the appropriate dispersion relation (with in general 
$\omega(-\kk) = - \omega(\kk)$) and $\HH$ is a symmetric function in its vector arguments 
which basically depends on the quadratic nonlinear operator ${\cal N}$.
Note the use of the Einstein's notation. We introduce 
\be
\AA(\kk,t) = \aa(\kk,t) e^{-i \omega_k t} \, ,
\ee
and obtain in the interaction representation
\be
{\partial a_j(\kk) \over \partial t} =
\varepsilon \int_{\RR^6} \HH_{jmn}^{\kk \pp \qq} a_m(\pp) a_n(\qq) 
e^{i \Omega_{k,pq}t} \delta_{k,pq} d\pp d\qq \, ,
\label{eq6}
\ee
where the Dirac delta function $\delta_{k,pq}=\delta(\kk-\pp-\qq)$ and 
$\Omega_{k,pq}=\omega_k -\omega_p -\omega_q$;  the time dependence in fields, $\aa$, is omitted for 
simplicity. Relation (\ref{eq6}) is the wave amplitude equation whose dependence in $\varepsilon$ means 
that weak nonlinearities will modify only slowly in time the wave amplitude. By nature, the problems
considered here (in MHD, electron and Hall MHD) involve mainly three-wave interaction processes as it is 
expected by the form of the wave amplitude equation. The exponentially oscillating term is essential for the 
asymptotic closure since we are interested in the long time statistical behavior for which the nonlinear transfer 
time is much greater than the wave period. In such a limit most of the nonlinear terms will be destroyed by 
random phase mixing and only a few of them -- called the resonance terms -- will survive. Before going to the 
statistical formalism, we note the following general properties that will be used
\ba
\HH_{jmn}^{\kk \pp \qq} &=& \big(\HH_{jmn}^{-\kk -\pp -\qq} \big)^* \, , \\
\HH_{jmn}^{\kk \pp \qq} &{\rm is}& {\rm \, symmetric \, in \,} (\pp, \qq) \, {\rm and \,} (m, n) \, , 
\label{proper} \\
\HH_{jmn}^{0 \pp \qq} &=& 0 \, ,
\ea
where, *, stands for the complex conjugate.

\subsection{Statistics and asymptotics}

We turn now to the statistical description, introduce the ensemble average $\langle ... \rangle$ and 
define the density tensor for homogeneous turbulence 
\be
q_{jj^\prime}(\kk^\prime) \delta(\kk+\kk^\prime) = \langle a_j(\kk)a_{j^\prime}(\kk^\prime) \rangle \, . 
\ee
We also assume that on average $\langle \uu(\xx,t) \rangle = 0$ which leads to the relation 
$\HH_{jmn}^{0 \pp \qq}=0$. From the nonlinear equation (\ref{eq6}), we find
\be
{\partial q_{jj^\prime} \delta(k+k^\prime)\over \partial t} = 
\left\langle a_{j^\prime}(\kk^\prime) {\partial a_j(\kk) \over \partial t} \right\rangle + 
\left\langle a_j(\kk) {\partial a_{j^\prime}(\kk^\prime) \over \partial t} \right\rangle = 
\label{eq8}
\ee
$$
\varepsilon \int_{\RR^6} \HH_{jmn}^{\kk \pp \qq} 
\big \langle a_m(\pp) a_n(\qq) a_{j^\prime}(\kk^\prime) \big\rangle 
e^{i \Omega_{k,pq}t} \delta_{k,pq} d\pp d\qq
$$
$$+$$
$$
\varepsilon \int_{\RR^6} \HH_{j^{\prime}mn}^{\kk^\prime \pp \qq} 
\big\langle a_m(\pp) a_n(\qq) a_j(\kk) \big\rangle 
e^{i \Omega_{k^\prime,pq}t} \delta_{k^\prime,pq} d\pp d\qq \, .
$$
A hierarchy of equations will clearly appear which gives for the third order moment equation
\be
{\partial  \big\langle a_j(\kk)a_{j^\prime}(\kk^\prime) a_{j^{\prime\prime}}(\kk^{\prime\prime}) \big\rangle 
\over \partial t} = 
\label{eq123}
\ee
$$
\varepsilon \int_{\RR^6} \HH_{jmn}^{\kk \pp \qq}  
\big\langle a_m(\pp) a_n(\qq) a_{j^\prime}(\kk^\prime) a_{j^{\prime\prime}}(\kk^{\prime\prime}) \big\rangle 
e^{i \Omega_{k,pq} t} \delta_{k,pq} d\pp d\qq
$$
$$
+ \, \varepsilon \int_{\RR^6} \Big\{ (\kk,j) \leftrightarrow (\kk',j') \Big\} d\pp d\qq
$$
$$
+ \, \varepsilon \int_{\RR^6} \Big\{ (\kk",j") \leftrightarrow (\kk',j') \Big\} d\pp d\qq \, ,
$$
where in the right hand side the second line means an interchange in the notations between two pairs with the first 
line as a reference, and the third line means also an interchange in the notations between two pairs with the second 
line as a reference. 
At this stage, we may write the fourth order moment in terms of a sum of the fourth order cumulant 
plus products of second order ones, but a natural closure arises for times asymptotically large 
\citep[see \eg,][]{Newell01,Naza11,Newell11}. In this case, several terms do not contribute at large times like, 
in particular, the fourth order cumulant which is not a resonant term. In other words, the 
nonlinear regeneration of third order moments depends essentially on products of second order
moments. The time scale separation imposes a condition of applicability of wave turbulence which 
has to be checked {\it in fine} \citep[see \eg,][]{Naza07}. After integration in time, we are left with
\be
\big\langle a_j(\kk)a_{j^\prime}(\kk^\prime) a_{j^{\prime\prime}}(\kk^{\prime\prime}) \big\rangle = 
\ee
$$
\varepsilon \int_{\RR^6} \HH_{jmn}^{\kk \pp \qq} 
\Big( \langle a_m(\pp) a_n(\qq) \rangle 
\langle a_{j^\prime}(\kk^\prime) a_{j^{\prime\prime}}(\kk^{\prime\prime}) \rangle
+
\langle a_m(\pp) a_{j^\prime}(\kk^\prime) \rangle
\langle a_n(\qq) a_{j^{\prime\prime}}(\kk^{\prime\prime}) \rangle
$$
$$
+
\langle a_m(\pp) a_{j^{\prime\prime}}(\kk^{\prime\prime}) \rangle
\langle a_n(\qq) a_{j^\prime}(\kk^\prime) \rangle \Big)
\Delta(\Omega_{k,pq}) \delta_{k,pq} d\pp d\qq
$$
$$
+ \, \varepsilon \int_{\RR^6} \Big\{ (\kk,j) \leftrightarrow (\kk',j') \Big\} d\pp d\qq
+ \, \varepsilon \int_{\RR^6} \Big\{ (\kk",j") \leftrightarrow (\kk',j') \Big\} d\pp d\qq \, ,
$$
where
\be
\Delta(\Omega_{k,pq}) = \int_0^{t\gg1/\omega} e^{i \Omega_{k,pq}t^\prime} dt^\prime 
=  {e^{i \Omega_{k,pq}t} - 1 \over i \Omega_{k,pq} } \, .
\ee
The same convention as in (\ref{eq123}) is used. After integration in wave vectors $\pp$ and $\qq$ and simplification, 
we get
\be
\big\langle a_j(\kk)a_{j^\prime}(\kk^\prime) a_{j^{\prime\prime}}(\kk^{\prime\prime}) \big\rangle = 
\ee
$$
\varepsilon \Delta(\Omega_{k k^\prime k^{\prime\prime}}) \delta_{kk^\prime k^{\prime\prime}}
$$
$$
\Big( \HH_{jmn}^{\kk -\kk^\prime -\kk^{\prime\prime}} 
q_{mj^\prime}(\kk^\prime) q_{n j^{\prime\prime}} (\kk^{\prime\prime})
+  \HH_{jmn}^{\kk -\kk^{\prime\prime} -\kk^\prime} 
q_{m j^{\prime\prime}} (\kk^{\prime\prime}) q_{n j^\prime} (\kk^\prime)
$$
$$
+ \HH_{j^\prime mn}^{\kk^\prime -\kk -\kk^{\prime\prime}}
q_{mj} (\kk) q_{nj^{\prime\prime}} (\kk^{\prime\prime})
+ \HH_{j^\prime mn}^{\kk^\prime -\kk^{\prime\prime} -\kk} 
q_{mj^{\prime\prime}} (\kk^{\prime\prime}) q_{nj} (\kk)
$$
$$
+ \HH_{j^{\prime\prime}mn}^{\kk^{\prime\prime} -\kk -\kk^\prime}  
q_{mj} (\kk)q_{nj^{\prime}} (\kk^\prime)
+ \HH_{j^{\prime\prime}mn}^{\kk^{\prime\prime} -\kk^\prime -\kk}  
q_{mj^{\prime}} (\kk^\prime) q_{nj} (\kk) \Big) \, .
$$
The symmetries (\ref{proper}) lead to
\be
\big\langle a_j(\kk)a_{j^\prime}(\kk^\prime) a_{j^{\prime\prime}}(\kk^{\prime\prime}) \big\rangle = 
\ee
$$
2 \varepsilon \Delta(\Omega_{k k^\prime k^{\prime\prime}}) \delta_{kk^\prime k^{\prime\prime}}
\Big( \HH_{jmn}^{\kk -\kk^\prime -\kk^{\prime\prime}} 
q_{mj^\prime}(\kk^\prime) q_{n j^{\prime\prime}} (\kk^{\prime\prime})
$$
$$
+ \HH_{j^\prime mn}^{\kk^\prime -\kk -\kk^{\prime\prime}}
q_{mj} (\kk) q_{nj^{\prime\prime}} (\kk^{\prime\prime})
+ \HH_{j^{\prime\prime}mn}^{\kk^{\prime\prime} -\kk -\kk^\prime}  
q_{mj} (\kk)q_{nj^{\prime}} (\kk^\prime) \Big) \, .
$$
The latter expression may be introduced into (\ref{eq8}). We take the long time limit (which introduces 
irreversibility) and find 
\be
\Delta(x) \to \pi \delta(x) + i {\cal P} (1/x) \, , 
\ee
with ${\cal P}$ the  principal value of the integral. We finally obtain the asymptotically exact wave kinetic 
equations
\be
{\partial q_{jj^\prime} (\kk) \over \partial t} = 
4 \pi \varepsilon^2 \int_{\RR^6} 
\delta_{k,pq} \delta(\Omega_{k,pq}) \HH_{jmn}^{\kk \pp \qq} 
\label{eqfin}
\ee
$$
\Big( \HH_{m r s}^{\pp -\qq -\kk} q_{rn}(\qq) q_{j^\prime s}(\kk)
+ \HH_{nrs}^{\qq -\pp \kk} q_{rm}(\pp) q_{j^\prime s} (\kk) 
+ \HH_{j^\prime rs}^{-\kk -\pp - \qq} q_{rm}(\pp) q_{sn}(\qq) \Big) d\pp d\qq \, .
$$
These general 3D wave kinetic equations are valid in principle for any situation where three-wave interaction 
processes are dominant; only the form of $\HH$ has to be adapted to the problem. Equation for the (total) 
energy is obtained by taking the trace of the tensor density, $q_{jj}(\kk)$, whereas other inviscid invariants 
are found by including non diagonal terms.

\subsection{Wave kinetic equations}

Equation (\ref{eqfin}) is the wave kinetic equation for the spectral tensor components. We see that the nonlinear 
transfer is based on a resonance mechanism since we need to satisfy the relations
\ba
\omega_k &=& \omega_p + \omega_q \, , \\
\kk &=& \pp+\qq \, . 
\ea
The solutions define the resonant manifolds which may have different forms according to the flow. 
For example in the limit of weakly compressible MHD, when the sound speed is much greater than 
the Alfv\'en speed, it is possible to find (for the shape of the resonant manifolds in the 3D k-space)
spheres or tilted planes for Fast-Fast-Alfv\'en and Fast-Fast-Slow wave interactions, and rays 
(a degenerescence of the resonant manifolds) for Fast-Fast-Fast wave interactions \citep{Galtier2001}. 
We also find planes perpendicular to the uniform magnetic field ${\bf B_0}$ for Slow-Slow-Slow, 
Slow-Slow-Alfv\'en, Slow-Alfv\'en-Alfv\'en or Alfv\'en-Alfv\'en-Alfv\'en wave interactions; it is a similar 
situation to incompressible MHD turbulence (since, at first order, slow waves have the same 
frequency as Alfv\'en waves) for which the resonant manifolds foliate the Fourier space. 

The representation of the resonant manifolds is always interesting since it gives an idea of how the 
spectral densities can be redistributed along (or transverse to) the mean magnetic field direction whose 
main effect is the nonlinear transfer reduction along its direction \citep{Bill96}. The previous finding was 
confirmed by a detailed analysis of wave compressible MHD turbulence \citep{kuznetsov01,chandran} 
in the small $\beta$ limit where the wave kinetic equations were derived as well as their exact power law 
solutions. The situation for electron and Hall MHD is more subtle and there is no simple picture for the 
resonant manifolds like in MHD. In this case, it is nevertheless important to check if the resonance condition
allows simple particular solutions in order to justify the domination of three-wave interaction processes over 
higher order (four-wave) processes. 

The form of the wave kinetic equations (\ref{eqfin}) is the most general one for a dispersive problem as 
whistler wave turbulence in electron MHD. The incompressible MHD system constitutes a unique example 
of pseudo-dispersive waves for which wave turbulence applies. In this particular case, some symmetries are 
lost and the principal value terms remain present. For that reason, incompressible MHD may be seen as a 
singular limit of incompressible Hall MHD \citep{Galtier06aa}.

\subsection{Finite flux solutions}
\label{K41solution}

The most spectacular result of the wave turbulence theory is its ability to provide exact finite flux solutions. 
These solutions are found after applying to the wave kinetic equations a conformal transform proposed first by 
\citet{Zakh65} for isotropic turbulence: it is the so-called Kolmogorov--Zakharov spectra. Because anisotropy is 
almost always present in magnetized plasmas, a bi-homogeneous conformal transform is more appropriate  
\citep{kuznetsov72}. This operation can only be performed if first one assumes axisymmetry. With this assumption
the wave kinetic equations (\ref{eqfin}) write
\be
{\partial {\tilde q}_{jj^\prime} (\kpe,\kpa) \over \partial t} = 
4 \pi \varepsilon^2 \int_{\RR^4} \delta_{k,pq} \delta(\Omega_{k,pq}) {\tilde \HH}_{jmn}^{\kk \pp \qq} 
\label{eqfin2}
\ee
$$
\Big( {\tilde \HH}_{m r s}^{\pp -\qq -\kk} {\tilde q}_{rn}(\qpe,\qpa) {\tilde q}_{j^\prime s}(\kpe,\kpe)
+ {\tilde \HH}_{nrs}^{\qq -\pp \kk} {\tilde q}_{rm}(\ppe,\ppa) {\tilde q}_{j^\prime s} (\kpe,\kpa) 
$$
$$
+ {\tilde \HH}_{j^\prime rs}^{-\kk -\pp - \qq} {\tilde q}_{rm}(\ppe,\ppa) {\tilde q}_{sn}(\qpe,\qpa) \Big)
d\ppe d\ppa d\qpe d\qpa \, ,
$$
where ${\tilde q}_{jj^\prime} (\kpe,\kpa) =q_{jj^\prime} (\kk)/ (2 \pi \kpe)$ and ${\tilde \HH}$ is a geometric operator.  
Note that we have performed an integration over the polar angle. Except for incompressible MHD for which the 
wave kinetic equations simplify thanks to the absence of nonlinear transfer along the parallel direction 
(along ${\bf B_0}$), in general we have to deal with a dynamics in the perpendicular and parallel 
directions with a relatively higher transfer transverse to ${\bf B_0}$ than along it. In this case, we may write the 
wave kinetic equations in the limit $\kpe \gg \kpa$. The system of integro-differential equations obtained is then 
sufficiently reduced to allow us to extract the exact power law solutions. We perform the conformal transformation 
to the equations for the invariant spectral densities (total energy, magnetic helicity...)
\be
\begin{array}{lll}
\ppe &\to & \kpe^2 / \ppe \, , \\[.2cm]
\qpe &\to & \kpe \qpe / \ppe \, , \\[.2cm]
\ppa &\to & \kpa^2 / \ppa \, , \\[.2cm]
\qpa &\to & \kpa \qpa / \ppa \, ,
\end{array}
\label{trans}
\ee
and we search for stationary solutions in the power law form $\kpe^{-n}\kpa^{-m}$. Basically two 
types of solutions are found: the fluxless solution, also called the thermodynamic equilibrium solution, 
which corresponds to the equipartition state for which the flux is zero, and the finite flux solution which 
is the most interesting one. During the last decades many papers have been devoted to the finding of 
these finite flux solutions for isotropic as well as anisotropic wave turbulence \citep{ZLF,Naza11}. 

\begin{figure}
\centerline{\psfig{file=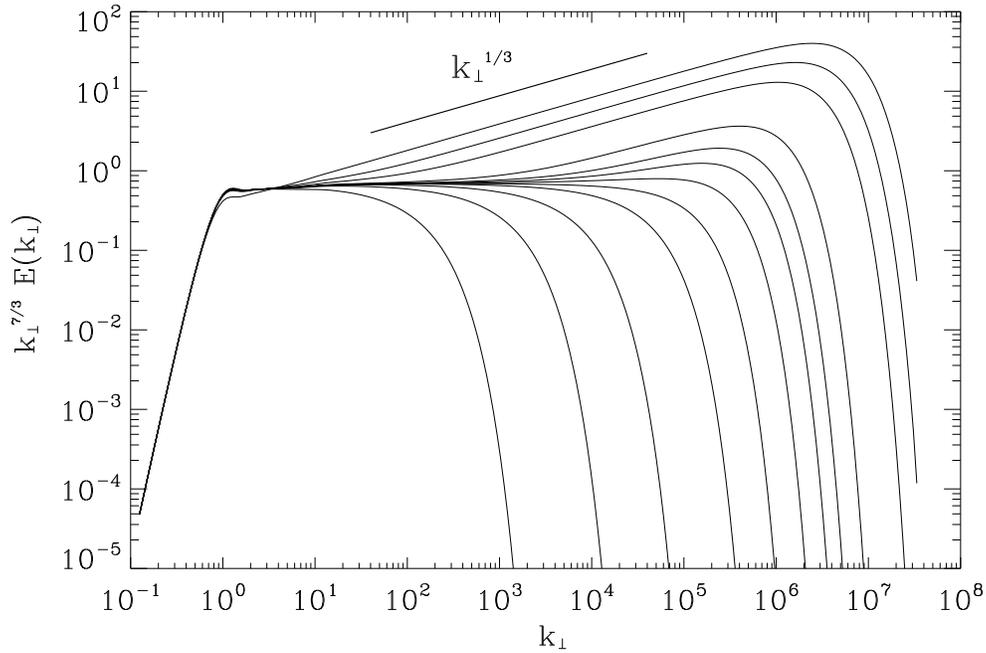,width=14cm}}
\caption{Time propagation (from left to right) of the compensated energy spectrum for incompressible Alfv\'en wave 
turbulence ($E=E^+=E^-$): an anomalous scaling is observed before the formation of the Kolmogorov--Zakharov 
spectrum.}
\label{fig3}
\end{figure}

Recently, and thanks to high numerical resolutions, a new challenge has appeared in wave turbulence: 
the study of incompressible Alfv\'en wave turbulence \citep{Galtier2000} reveals that the development of the 
finite flux solution of the wave kinetic equations is preceded by a front characterized by a significantly steeper 
scaling law. An illustration is given in Fig. \ref{fig3} for the balance case ($E=E^+=E^-$): the temporal evolution 
of the energy spectrum reveals the formation of a $\kpe^{-7/3}$ front before the establishment of the 
Kolmogorov-Zakharov solution in $\kpe^{-2}$. This finding is in contradiction with the theory 
proposed by \citet{Falko} on the nonlinear front propagation where the authors claimed that the 
Kolmogorov--Zakharov spectrum should be formed right behind the propagating front. The same 
observation was also made for the inverse cascade in the nonlinear Schr\"odinger equation \citep{Lacaze} 
and a detailed analysis was given by \citet{colm} in the simplified case of strongly local interactions for which 
the usual wave kinetic equations become simple PDEs \citep[for MHD, see,][]{GB10}. In spite of these attempts, 
the anomalous scaling is still an open problem without clear physical and proper mathematical answer. 

\section{Main results and predictions}
\label{4}

\subsection{Alfv\'en wave turbulence}
\label{galtsol}

We start to summarize the results of wave turbulence in the incompressible MHD case for which the 
wave kinetic equations are singular in the sense that the principal value terms remain except for the 
Els\"asser energies $E^\pm$. As it was explained, the origin of this particularity is the pseudo-dispersive 
nature of Alfv\'en waves which are a unique case where wave turbulence theory applies \citep[see also][]{ZS70}.
In the limit of strongly anisotropic turbulence for which $\kpe \gg \kpa$, the wave kinetic equations (\ref{eqfin2}) 
simplify. Then, the Els\"asser energy spectra of the transverse fluctuations (shear-Alfv\'en waves) satisfy the 
asymptotic integro-differential equations 
\be
{\partial E^\pm (\kpe,\kpa) \over \partial t} = 
\label{shearEq}
\ee
$$
{\pi \varepsilon^2 \over B_0} \iint_{\Delta} \cos^2\phi \sin\theta \, \frac{\kpe}{\qpe}
E^\mp(\qpe,0) \left[\kpe E^\pm(\ppe,\kpa) - \ppe E^\pm(\kpe,\kpa)\right] d\ppe d\qpe , 
$$
where $\Delta$ defines the domain of integration on which ${\bf \kpe} = {\bf \ppe} + {\bf \qpe}$, $\phi$ is the
angle between ${\bf \kpe}$ and ${\bf \ppe}$ and $\theta$ the angle between ${\bf \kpe}$ and ${\bf \qpe}$. 
A fundamental property of Alfv\'en wave turbulence appears on equations (\ref{shearEq}): the evolution of the 
spectra $E^\pm$ is always mediated by interaction with the slow mode $\qpa=0$. This is a direct consequence 
of the resonance conditions discussed in Section \ref{resona}. Then, the parallel wavenumber dependence does 
not affect the nonlinear dynamics: in other words, there is no transfer along $\bf {B_0}$ and $\kpa$ can be treated 
as an external parameter. Therefore, we introduce the notation $E^\pm(\kpe,\kpa) = E^\pm(\kpe) f_\pm (\kpa)$ 
where $f_\pm$ are arbitrary functions of $\kpa$ given by the initial conditions and with the assumption $f_\pm(0)=1$ 
we obtain 
\be
{\partial E^\pm (\kpe) \over \partial t} = 
\label{master}
\ee
$$
{\pi \varepsilon^2 \over B_0} \iint_{\Delta} \cos^2\phi \sin\theta \, \frac{\kpe}{\qpe}
E^\mp(\qpe) \left[\kpe E^\pm(\ppe) - \ppe E^\pm(\kpe)\right] d\ppe d\qpe ,
$$
which describes the transverse dynamics. The exact finite flux solutions of equations (\ref{master}) for the 
stationary energy spectra are 
\be
E^\pm(\kpe) \sim \kpe^{n_\pm} 
\ee
with 
\be
n_+ + n_- = -4 \, .
\label{KZs}
\ee
Additionally, the power law indices must satisfy the condition of locality (\ie the condition on the power law 
solutions for which the wave kinetic equations remain finite) 
\be
-3 < n_\pm < -1 \, .
\label{36}
\ee
It is important to recall that wave turbulence is an asymptotic theory which must satisfy a condition of application. 
In our case the transfer time must be significantly larger than the Alfv\'en time which means in terms of wavenumbers 
that $\kpa \gg \varepsilon^2 \kpe$ (and that at small transverse scales turbulence becomes strong). 
At this point a comment has to made about the Kolmogorov--Zakharov 
solutions (\ref{KZs}). These solutions imply the contribution of the wavevector $\qq$ which is by nature a slow 
mode ($\qpa=0$) whereas the other contributions $\kk$ and $\pp$ imply wave modes ($\kpa=\ppa>0$). 
In Fig. \ref{fig3} it is assumed that $E^+=E^-$ and that the dynamics of the slow mode is given by the wave mode
hence the stationary solution $n_+=n_-=2$. However, it may happen that the slow mode has its own dynamics
which can belong to strong turbulence. Since relation (\ref{KZs}) is exact, if the wave mode energy has a $-5/3$ 
scaling then the wave turbulence spectrum should be $-7/3$. Although this discussion was given in the original 
paper \citep{Galtier2000} it was not considered seriously until the most recent direct numerical simulations 
\citep{Bigot08b,Bigot11} reveal that this situation may happen. We will come back to this point in the conclusion.

The locality of interactions (in Fourier space) is an important issue in MHD turbulence. According to some 
recent works in isotropic turbulence, nonlinear interactions seem to be more non local in MHD than in a 
pure hydrodynamics in the sense that the transfer of energy from the velocity field to the magnetic field 
may be a highly non local process in Fourier space \citep{Alexakis05}. The situation is different when 
we deal with anisotropic turbulence: in this case interactions (between perpendicular wavevectors) 
are mainly local \citep{Alexakis07}. In wave turbulence, the condition (\ref{36}) has to be satisfied to 
validate the exact power law solutions and avoid any divergence of integrals in the wave kinetic equations 
due to nonlocal contributions. In practice, numerical simulations of the wave kinetic equations have clearly 
shown that the solutions (\ref{36}) are attractive \citep{Galtier2000}.

Recently, it was realized that the wave kinetic equations found in the anisotropic limit may be 
recovered without the wavenumber condition $\kpe \gg \kpa$ if initially only the shear-Alfv\'en waves were 
considered \citep{Galtier06b}. Therefore, the finite flux solution may be extended to the entire wavenumber 
space which renders its detection easier. This idea was tested successfully against numerical simulations
by \citet{Perez08} who found a spectral signatures of Alfv\'en wave turbulence.

\subsection{Compressible MHD}

We turn now to the compressible regime for which two limits have been analyzed. The first case (case I) 
is the one for which the plasma (thermal) pressure is assumed to be small as compared to the magnetic 
pressure (small $\beta$ limit) and where three-wave interaction processes that do not involve slow waves 
are neglected \citep{kuznetsov01}. In the second case (case II) for which we still have $\beta \ll 1$, slow 
waves are neglected and a constant density is imposed \citep{chandran}. In both situations the general 
finite flux solutions are not obvious to express. 

In case I, when only interactions between Alfv\'en and slow waves are kept, a wave energy spectrum 
in $\sim \kpe^{-2} \kpa^{-5/2}$ is found which corresponds to a (finite) constant energy flux 
solution. It is claimed that the addition of the interactions with the fast waves will lead to the same solution 
since the dynamics tends to produce strongly anisotropic distributions of the waves concentrated in 
k-space within a narrow-angle cone in the ${\bf B_0}$ direction ($\kpe \ll \kpa$). Under these conditions, the 
fast waves coincide with the Alfv\'en waves. 

In case II, the general wave kinetic equations do not allow us to find exact power law solutions. However, 
when only Fast-Fast-Fast interactions are kept it is possible to find a finite flux solution for the 
fast wave 1D energy spectrum which scales as $\sim f(\theta) k^{-3/2}$, where $f(\theta)$ is an arbitrary 
function of the angle $\theta$ between the wave vector $\kk$ and the uniform magnetic field ${\bf B_0}$.
Numerical simulations of the general wave kinetic equations are made to find the behavior according to 
the angle $\theta$. A solution close to $\kpe^{-2}$ is found for the Alfv\'en wave 2D energy 
spectrum (for different fixed $\kpa$) when $\kpe \gg \kpa$. The $k$--spectra plotted at $\theta =45^o$ 
reveal a fast wave modal energy spectrum in $\sim k^{-3/2}$ and an steeper Alfv\'en wave spectrum, while 
for a small angle ($\theta =7.1^o$) both spectra follow the same scaling law steeper than $k^{-3/2}$ 
\citep{chandran}.

\subsection{Whistler wave turbulence}

The electron MHD equations in the presence of a strong uniform magnetic field $B_0$ exhibit 
dispersive whistler waves. The wave turbulence regime was analyzed in the incompressible 
case by \citet{Galtier03,Galtier05a} who derived the wave kinetic equations by using a complex 
helicity decomposition. The strong anisotropic ($\kpe \gg \kpa$) finite flux solutions correspond to 
\be
E(\kpe,\kpa) \sim \kpe^{-5/2} \kpa^{-1/2} 
\ee
for the magnetic energy spectrum, and 
\be
H(\kpe,\kpa) \sim \kpe^{-7/2} \kpa^{-1/2} 
\ee
for the magnetic helicity spectrum. As for the other cases presented above, a direct cascade was 
found for the energy. In particular it was shown that contrary to MHD, the wave kinetic equations which 
involve three-wave interaction processes are characterized by a nonlinear transfer that decreases linearly 
with $\kpa$; for $\kpa=0$, the transfer is exactly null. Thus the 2D modes (or slow modes) decouple from 
the three-dimensional whistler waves. Such a decoupling is found in a variety of problems like rotating 
turbulence \citep{Galtier2003,sagaut}.

\subsection{Hall MHD}

The last example exposed in this chapter is the Hall MHD case which incorporates both the standard MHD 
and electron MHD limits. This system is much heavier to analyze in the regime of wave 
turbulence and it is only recently that a theory has been proposed \citep{Galtier06aa}. The general theory 
emphasizes the fact that the large scale limit of standard MHD becomes singular with the apparition of 
a new type of terms, the principal value terms. Of course, the large scale and small scale limits tend to 
the appropriate theories (MHD and electron MHD wave turbulence), but in addition it is possible to describe 
the connection between them at intermediate scales (scales of the order of the ion inertial length $d_i$). For 
example, moderate anisotropy is predicted at these intermediate scales whereas it is much stronger for other 
scales. It is also interesting to note that the small scale limit gives a system of equations richer than the 
pure electron MHD system \citep{Galtier03} with the possibility to describe the sub-dominant kinetic 
energy dynamics. An exact power law solution for the kinetic energy spectrum is found for ion cyclotron 
wave turbulence which scales as
\be 
E(\kpe,\kpa) \sim \kpe^{-5/2} \kpa^{-1/2} \, .
\ee
To date, the wave kinetic equations of Hall MHD have not been simulated numerically even in their simplified 
form (when helicity terms are discarded). It is an essential step to understand much better the dynamics at 
intermediate scales.

\section{Conclusion and perspectives}
\label{5}

\subsection{Observations}

\begin{figure}
\centerline{\psfig{file=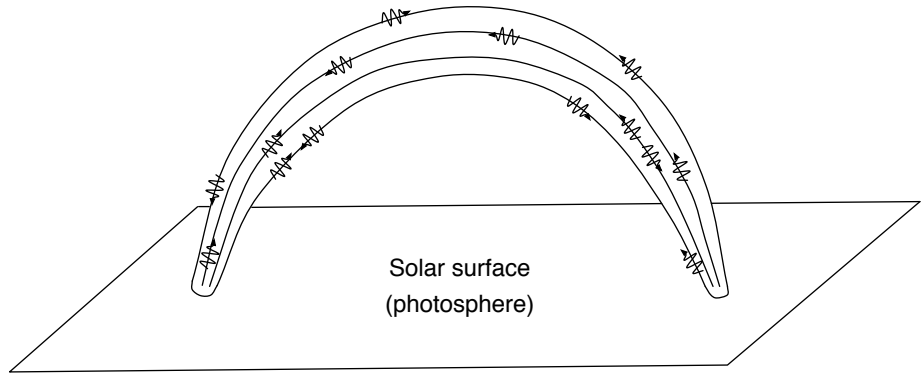,width=12.5cm}}
\caption{Schematic view of Alfv\'en wave turbulence on the Sun: coronal loops act as a resonant cavity for 
Alfv\'en wave packets which are transmitted from the deeper layers of the solar atmosphere.}
\label{fig4}
\end{figure}

Waves and turbulence are two fundamental ingredients of many space plasmas. The most spectacular 
illustration of such characteristics is probably given by the observations of the Sun's atmosphere with the 
orbital solar observatory Hinode/JAXA and more recently by SDO/NASA launched in 2010. For the first time, 
detection of Alfv\'en waves is made through the small oscillations of many thin structures called threads
(see Fig. \ref{fig4}). In the meantime the highly dynamical nature of coronal loops is revealed by 
non-thermal velocities detected with spectrometers. These findings are considered as a remarkable step 
in our understanding of the solar coronal dynamics. Nowadays, it is believed that Alfv\'en waves turbulence 
is a promising model to understand the heating of the solar corona \citep{vanB}. 

The interplanetary medium is another example of magnetized plasma where waves and turbulence 
are detected. In this framework, the origin of the so-called "dissipative range", \ie the extension of the 
turbulent inertial range beyond a fraction of hertz, is currently one of the main issue discussed in the 
community. Although a final answer is not given yet, wave turbulence is a promising regime to 
understand the inner solar wind dynamics in the sense that it gives exact results in regards to the 
possible multiscale behavior of magnetized plasmas as well as the intensity of the anisotropic transfer 
between modes. 

The main feature of magnetized plasmas in the regime of wave turbulence is the omnipresence of 
anisotropy and the possibility to have different spectral scaling laws according to the space direction. 
To achieve a proper comparison between observational data and theoretical predictions, not only 
{\it in situ} measurements are necessary, but multipoints data have also to be accessible. It is at this 
price that the true nature of magnetized plasmas will be revealed. The magnetosphere is the first 
medium where it is possible to perform such a comparison thanks to the Cluster/ESA mission 
\citep{Sahraoui06}. In the case of the Jupiter's magnetosphere the large scale magnetic field is relatively 
strong. Actually, the first indirect evidence of Alfv\'en wave turbulence was reported in \citet{Saur} by 
using five years set of Galileo/NASA spacecraft magnetic field data. (Since the work was based on a 
one point analysis a model was used to differentiate the perpendicular and parallel spectra.)

Perhaps the first direct evidence of Alfv\'en wave turbulence comes from solar (photospheric) magnetic field 
observations above active regions. Using data from SOHO/ESA, \citet{abramenko} observed that the magnetic 
energy spectra reduced to its radial component (along ${\bf B_0}$) have a transverse scaling significantly different 
from the Kolmogorov one with power laws generally between $-2$ and $-2.3$. Since the magnetic field can only 
be detected in a thin layer (the temperature is too high in the chromosphere and in the corona to measure the 
magnetic field with the Zeeman effect), the magnetic spectra can be interpreted as a measure at a given 
$\kpa>0$ (in practice a small average is made in $\kpa$ since the data comes from slightly different altitudes). 
As we know from the Alfv\'en wave turbulence theory the parallel fluctuations follow the same dynamics as the 
perpendicular (shear-Alfv\'en) fluctuations and the addition of the power law indices of the slow and wave mode 
spectra is $-4$. Then, these unexpected spectra may be seen as the direct manifestation of Alfv\'en wave 
turbulence if the slow mode contribution scales like Kolmogorov (see the discussion on Section \ref{galtsol}). 
If this interpretation is correct it is probably the most important achievement of the Alfv\'en wave turbulence 
theory. 

\subsection{Simulations}

Numerical simulation is currently the main tool to improve our knowledge on wave turbulence in 
magnetized plasmas since we are still limited by the observational (single point) data. Two types 
of simulations are available: the simulation of the wave kinetic equations and the direct 
numerical simulation of the original MHD-type equations. In the former case, it is a way to find for 
instance the spectral scaling laws when the wave kinetic equations are too complex to provide us 
the exact solutions after application of the usual conformal transform. An example is given by 
compressible MHD: the numerical simulations revealed a variation of the power law energy 
spectrum with the angle between $\kk$ and ${\bf B_0}$ \citep{chandran}. Simulations may also be 
useful to investigate wave turbulence when an external forcing is applied like in incompressible 
MHD \citep{Galtier08c}. 

For direct numerical simulation, the challenge is slightly different: indeed, in this case the main 
goal is the measure of the transition between strong and wave turbulence, and thus between isotropic 
and anisotropic turbulence. The former regime has been extensively studied since more than 
three decades whereas the latter is still a young subject. The main topic of such a simulation is also 
to find general properties that could help us to understand the single point measurements made in 
natural plasmas. We arrive here at the heart of current issues in wave turbulence. One of the most 
important points emphasized by recent direct numerical simulations in incompressible MHD is the 
coexistence of wave and strong turbulence \citep{Bigot08b}. This characteristic should not be a 
surprise since basically wave turbulence is a perturbative theory which must satisfy conditions of 
applicability. In this case, the slow mode ($\qpa=0$) may evolve differently from the wave modes 
($\kpa,\ppa>0$) since the former case may be characterized by strong turbulence and the latter by wave 
turbulence. This distinction is fundamental and not really taken into account in the community: Alfv\'en 
wave turbulence will not be fully revealed in space as well as simulated plasmas as long as 
strong and wave turbulence will not be separated. This point is illustrated with direct numerical simulations 
\citep{Bigot08b,Bigot11} where steeper energy spectra may be found for the wave modes (between 
$-2$ and $-7/3$) compared to the slow mode (down to $-5/3$). The result seems to depend on the relative 
intensity of $B_0$: a strong magnetic background tends to increase the energy of the slow mode which, 
then, may evolve independently (the critical value of $B_0$ seems to be around $10$ times the \rms 
fluctuations). It is also found that the wave turbulence spectra are generally visible at a fixed $\kpa>0$ and 
may be hidden in a $-5/3$ scaling if a summation over $\kpa$ is performed (this result could be due to the 
limited space resolution). The same investigation clearly shows that -- as expected-- an equipartition between 
the kinetic and magnetic energies is reached in the wave modes whereas the magnetic energy is dominant in 
the slow mode. Note that this Alfv\'en wave turbulence regime was also numerically found independently 
by \citet{Perez08}.

\subsection{Open questions}

In the light of such recent results the following new questions may be addressed:

\medskip

$\bullet$ Does the $-7/3$ spectrum more universal than $-2$ in Alfv\'en wave turbulence ?

\smallskip

$\bullet$ Is the slow mode ($\qpa=0$) intermittent ? and then does it produce intermittency in the slave 
wave modes ($\kpa,\ppa>0$) ?

\smallskip

$\bullet$ Can we find direct signatures of Alfv\'en wave turbulence in solar magnetic loops with 
three-dimensional magnetic field data ?

\smallskip

$\bullet$ Does Alfv\'en wave turbulence the universal regime for stellar magnetic loops ?

\smallskip

$\bullet$ Does the $-5/3$ energy spectrum observed in the solar wind correspond to a bias ? and 
then what is the wave vector energy spectrum ?

\smallskip

$\bullet$ Is the small energy ratio $E^u/E^b \sim 0.5$ found in the local solar wind \citep{bruno} due 
essentially to the slow mode contribution in which case the wave modes may follow equipartition ?

\smallskip

$\bullet$ Are the absence of intermittency and the $-2.5$ magnetic spectrum observed in the solar wind 
dispersive regime \citep{kiyani} the signatures of whistler wave turbulence ?

\medskip

Future missions like {\it Solar Orbiter} from the European Space Agency (scheduled for 2017) will certainly 
help us to answer these questions and make significant advances on plasma turbulence.


\end{document}